\documentclass{aipproc}
\usepackage{hyperref}
\usepackage{multicol}
\newcommand{\com}[1]{}

\layoutstyle{6x9}
\date{\today}

\begin{document}
\bibliographystyle{aipproc}
\title{NA49 results on hadron production:\\
indications of the onset of deconfinement ?}

\classification{}
\keywords{}

\author{Benjamin Lungwitz for the NA49 collaboration$^0$\footnotetext{presented at the XXXVth International Symposium on Multiparticle Dynamics 2005.}}{
address={Institut für Kernphysik, Johann-Wolfgang Goethe Universit\"at Frankfurt, \\Max-von-Laue-Str. 1, 60438 Frankfurt am Main}
}

\maketitle

%\vspace{0.5cm}
\noindent
C.~Alt$^{9}$, T.~Anticic$^{21}$, B.~Baatar$^{8}$,D.~Barna$^{4}$,
J.~Bartke$^{6}$, L.~Betev$^{10}$, H.~Bia{\l}\-kowska$^{19}$,
C.~Blume$^{9}$,  B.~Boimska$^{19}$, M.~Botje$^{1}$,
J.~Bracinik$^{3}$, R.~Bramm$^{9}$, P.~Bun\v{c}i\'{c}$^{10}$,
V.~Cerny$^{3}$, P.~Christakoglou$^{2}$, O.~Chvala$^{14}$,
J.G.~Cramer$^{16}$, P.~Csat\'{o}$^{4}$, P.~Dinkelaker$^{9}$,
V.~Eckardt$^{13}$, 
%H.G.~Fischer$^{10}$,
D.~Flierl$^{9}$, Z.~Fodor$^{4}$, P.~Foka$^{7}$,
V.~Friese$^{7}$, J.~G\'{a}l$^{4}$,
M.~Ga\'zdzicki$^{9,11}$, V.~Genchev$^{18}$, G.~Georgopoulos$^{2}$, 
E.~G{\l}adysz$^{6}$, K.~Grebieszkow$^{20}$,
S.~Hegyi$^{4}$, C.~H\"{o}hne$^{7}$, 
K.~Kadija$^{21}$, A.~Karev$^{13}$, M.~Kliemant$^{9}$, S.~Kniege$^{9}$,
V.I.~Kolesnikov$^{8}$, E.~Kornas$^{6}$, 
R.~Korus$^{11}$, M.~Kowalski$^{6}$, 
I.~Kraus$^{7}$, M.~Kreps$^{3}$, A.~Laszlo$^{4}$, M.~van~Leeuwen$^{1}$, 
P.~L\'{e}vai$^{4}$, L.~Litov$^{17}$, B.~Lungwitz$^{9}$,
M.~Makariev$^{17}$, A.I.~Malakhov$^{8}$, 
M.~Mateev$^{17}$, G.L.~Melkumov$^{8}$, A.~Mischke$^{1}$, M.~Mitrovski$^{9}$, 
J.~Moln\'{a}r$^{4}$, St.~Mr\'owczy\'nski$^{11}$, V.~Nicolic$^{21}$,
G.~P\'{a}lla$^{4}$, A.D.~Panagiotou$^{2}$, D.~Panayotov$^{17}$,
A.~Petridis$^{2}$, M.~Pikna$^{3}$, D.~Prindle$^{16}$,
F.~P\"{u}hlhofer$^{12}$, R.~Renfordt$^{9}$,
C.~Roland$^{5}$, G.~Roland$^{5}$, 
M. Rybczy\'nski$^{11}$, A.~Rybicki$^{6,10}$,
A.~Sandoval$^{7}$, N.~Schmitz$^{13}$, T.~Schuster$^{9}$, P.~Seyboth$^{13}$,
F.~Sikl\'{e}r$^{4}$, B.~Sitar$^{3}$, E.~Skrzypczak$^{20}$,
G.~Stefanek$^{11}$, R.~Stock$^{9}$, C.~Strabel$^{9}$, H.~Str\"{o}bele$^{9}$, T.~Susa$^{21}$,
I.~Szentp\'{e}tery$^{4}$, J.~Sziklai$^{4}$, P.~Szymanski$^{10,19}$,
V.~Trubnikov$^{19}$, D.~Varga$^{4,10}$, M.~Vassiliou$^{2}$,
G.I.~Veres$^{4,5}$, G.~Vesztergombi$^{4}$,
%S.~Wenig$^{10}$,
D.~Vrani\'{c}$^{7}$, A.~Wetzler$^{9}$,
Z.~W{\l}odarczyk$^{11}$ I.K.~Yoo$^{15}$, J.~Zim\'{a}nyi$^{4}$

\vspace{0.5cm}
\noindent
$^{1}$NIKHEF, Amsterdam, Netherlands. \\
$^{2}$Department of Physics, University of Athens, Athens, Greece.\\
$^{3}$Comenius University, Bratislava, Slovakia.\\
$^{4}$KFKI Research Institute for Particle and Nuclear Physics, Budapest, Hungary.\\
$^{5}$MIT, Cambridge, USA.\\
$^{6}$Institute of Nuclear Physics, Cracow, Poland.\\
$^{7}$Gesellschaft f\"{u}r Schwerionenforschung (GSI), Darmstadt, Germany.\\
$^{8}$Joint Institute for Nuclear Research, Dubna, Russia.\\
$^{9}$Fachbereich Physik der Universit\"{a}t, Frankfurt, Germany.\\
$^{10}$CERN, Geneva, Switzerland.\\
$^{11}$Institute of Physics \'Swi{\,e}tokrzyska Academy, Kielce, Poland.\\
$^{12}$Fachbereich Physik der Universit\"{a}t, Marburg, Germany.\\
$^{13}$Max-Planck-Institut f\"{u}r Physik, Munich, Germany.\\
$^{14}$Institute of Particle and Nuclear Physics, Charles University, Prague, Czech Republic.\\
$^{15}$Department of Physics, Pusan National University, Pusan, Republic of Korea.\\
$^{16}$Nuclear Physics Laboratory, University of Washington, Seattle, WA, USA.\\
$^{17}$Atomic Physics Department, Sofia University St. Kliment Ohridski, Sofia, Bulgaria.\\ 
$^{18}$Institute for Nuclear Research and Nuclear Energy, Sofia, Bulgaria.\\ 
$^{19}$Institute for Nuclear Studies, Warsaw, Poland.\\
$^{20}$Institute for Experimental Physics, University of Warsaw, Warsaw, Poland.\\
$^{21}$Rudjer Boskovic Institute, Zagreb, Croatia.\\

\noindent\textbf{Abstract.} The NA49 experiment at the CERN SPS measured the energy and system size dependence of particle production in A+A collisions. 
A change of the energy
dependence of several hadron production properties at low SPS energies is observed which suggests a scenario requiring the 
onset of deconfinement.

%\begin{multicols}{2}

\section{Introduction}
In heavy ion interactions at sufficiently high collision energy the creation of a deconfined state of matter, the quark gluon plasma (QGP), is expected
\cite{Karsch:2000kv, Fodor:2004nz}.
At RHIC and top SPS energies the energy density of the fireball in the early stage was estimated to be large enough to create QGP. In contrast at low AGS
energies the maximum energy density is probably too low for the creation of deconfined matter.
In order to look for the onset of deconfinement the NA49 experiment at the CERN SPS studied the hadronic final state of 
Pb+Pb collisions in the energy range $20A$ - $158A$ GeV. \\
In addition to the energy scan program, NA49 also measured the dependence of hadron production properties on the size of the colliding nuclei and the
centrality of collisions.

\section{The NA49 Experiment}
\com{
\begin{figure}[htp]
\includegraphics[height=5cm]{aufbau2}
\caption{\label{NA49exp}Setup of the NA49 experiment.}
\end{figure}}
The NA49 detector system~\cite{Afanasev:1999iu}
% (figure~\ref{NA49exp}) 
is a large acceptance fixed target hadron spectrometer. Its main devices are four large 
time projection chambers (TPCs).
Two of them, called vertex TPCs, are located in two superconducting dipole magnets. The other two TPCs are installed behind the magnets left and
right of the beam line allowing precise particle tracking in the high density region of heavy ion collisions.\\
The accuracy of momentum determination is $\Delta p/p^2 \approx (0.3 - 7) \cdot 10^{-4}$ (GeV/c)$^{-1}$.

The measurement of the energy loss $dE/dx$ allows identification of pions, protons, kaons and electrons for momenta $p > 4$ GeV/c.
The NA49 experiment is equipped with two time of flight walls which, together with the energy loss measurement, allow a good separation 
of the particle species at midrapidity.
Decaying particles can be identified with a good precision via the reconstruction of the decay vertex and / or the invariant mass method.

The downstream veto calorimeter allows a determination of the centrality of a collision by measuring the energy in the projectile spectator region.
This measure of centrality is independent of the multiplicity of produced particles which is important in a study of multiplicity
fluctuations.

A large variety of different colliding systems was studied. Beams of p and Pb are available at the CERN SPS directly, C and Si beams 
were produced via fragmentation of the primary Pb beam. Targets of liquid hydrogen or solid foils of different materials were used.

In this paper results of the energy scan, which covers Pb+Pb collisions at $20A$, $30A$, $40A$, $80A$ and $158A$ GeV, and results of the
system size scan at $40A$ and $158A$ GeV are shown. The data should be considered preliminary unless a reference to the corresponding publication
is given.

\section{Results}
\noindent\textbf{Pions:}\\
Most of the produced particles are pions and thus they carry most of the entropy produced in the collision.
\begin{figure}[htp]
\includegraphics[height=5cm]{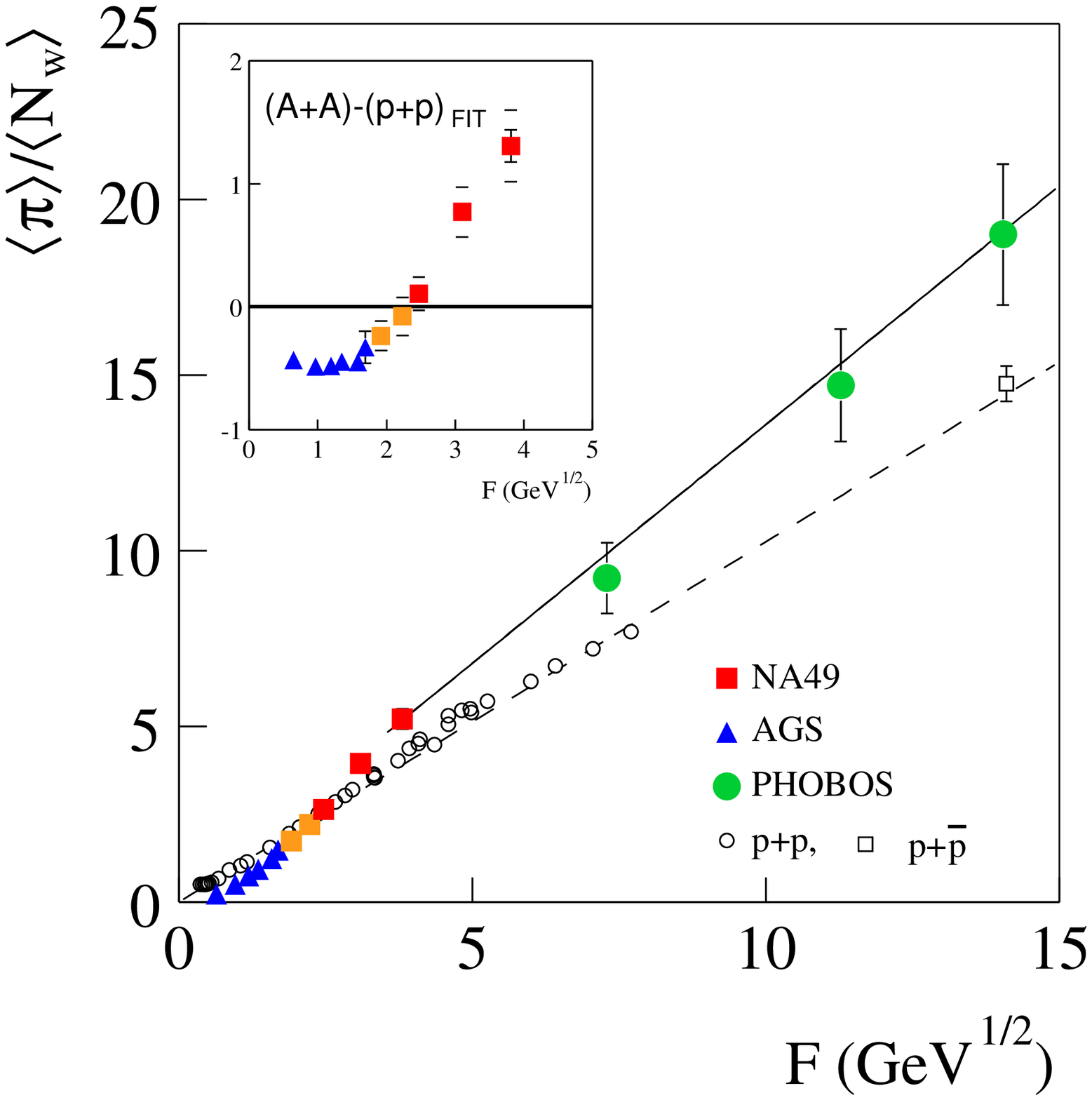}
\includegraphics[height=5cm]{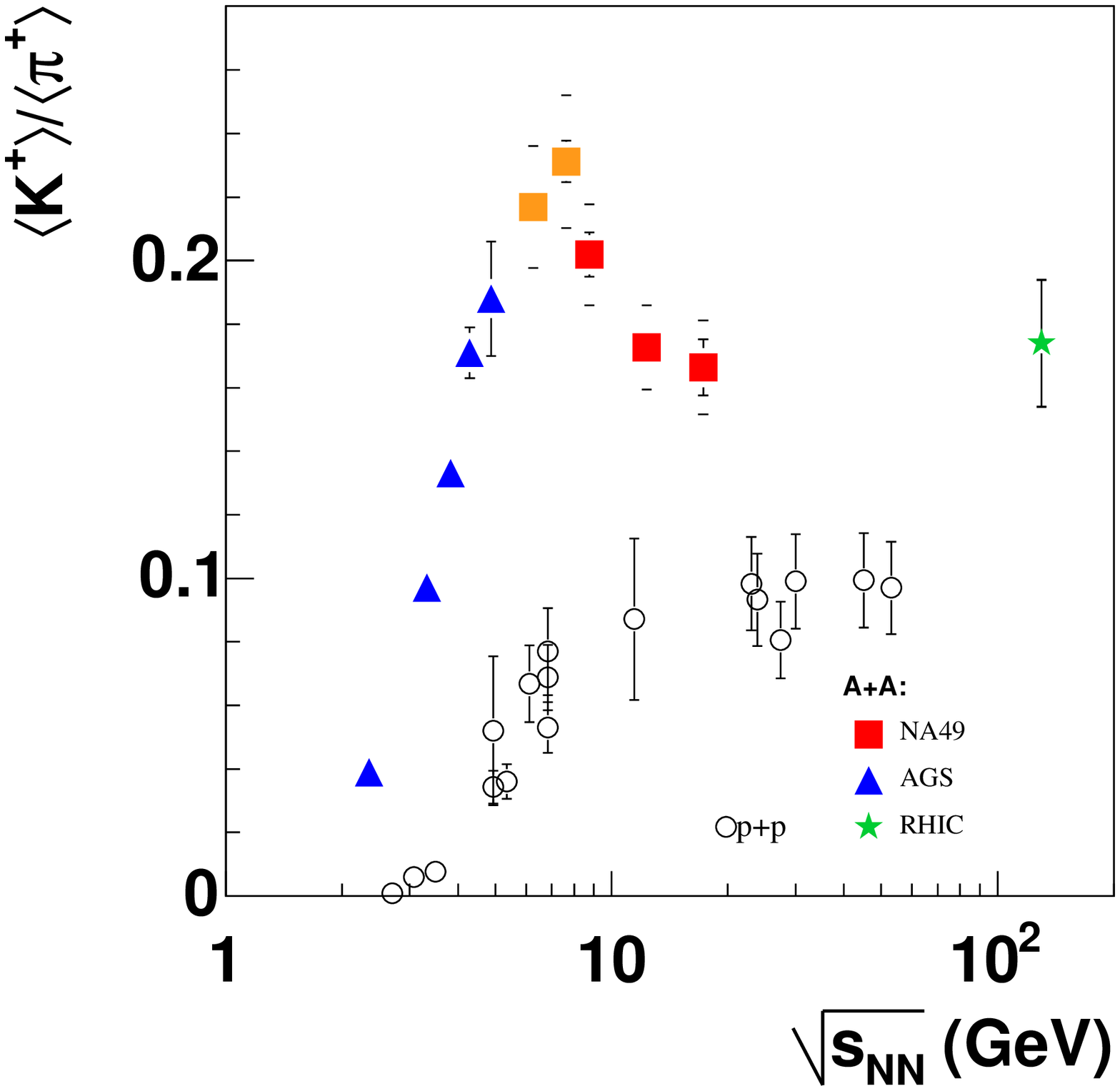}
\includegraphics[height=5cm]{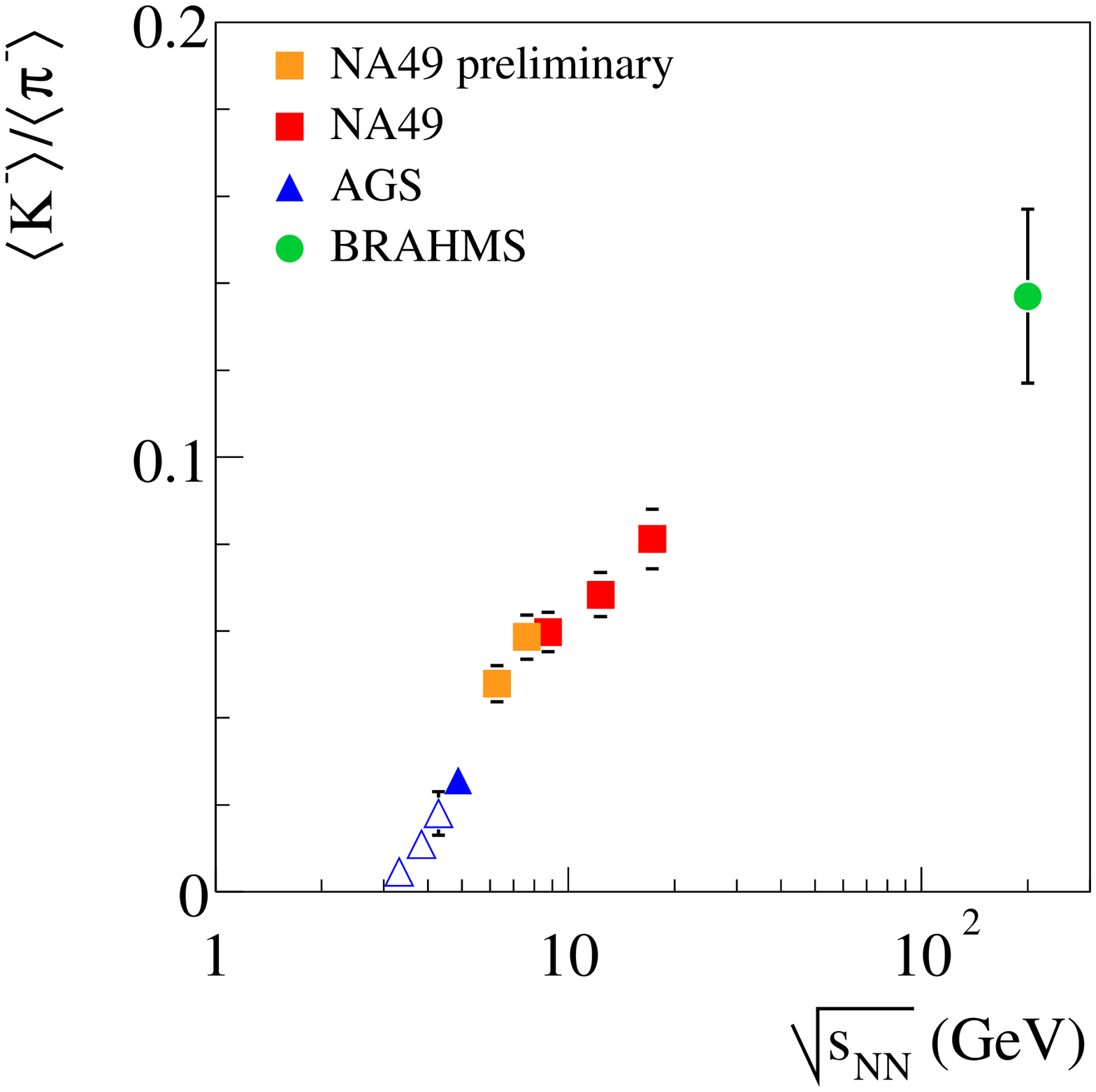}
\caption{\label{pi_ed}Energy dependence of pion multiplicity $\left<\pi\right>$ to the nubmer of wounded nucleons 
$\left<N_W\right>$~\cite{Gazdzicki:2004ef} (left),
of the $\left<K^+\right> / \left<\pi^+\right>$ ratio (middle) and the $\left<K^-\right> / \left<\pi^-\right>$ ratio~\cite{Gazdzicki:2004ef} (right) in Pb+Pb~(Au+Au) (full symbols) and 
p+p($\bar{\rm{p}}$) (open symbols) collisions.}
\end{figure}

Figure~\ref{pi_ed} shows the pion yield per wounded nucleon in central Pb+Pb (Au+Au) collisions and p+p interactions as a function of the
Fermi variable \\$F=\left( \sqrt{s_{NN}}-2 m_N \right) ^{3/4} / \left( \sqrt{s_{NN}} \right) ^{1/4} \approx \sqrt{\sqrt{s_{NN}}}$. 
The ratio $\left<\pi\right>/\left<N_W\right>$ rises linearly with $F$ for p+p interactions. For heavy ion collisions at low energies $\left<\pi\right>/\left<N_W\right>$ is smaller than for p+p 
by a constant amount which may be attributed~\cite{Gazdzicki:1998vd} to pion absorption in the hadronic medium.
At low SPS energies the pion multiplicity in Pb+Pb collisions starts to increase faster with energy than in p+p interactions.
This enhancement of pion production may indicate an increase of the number of degrees of freedom in the QGP phase 
and suggests the onset of deconfinement at low SPS energies.
\begin{figure}[htp]
\includegraphics[height=5cm]{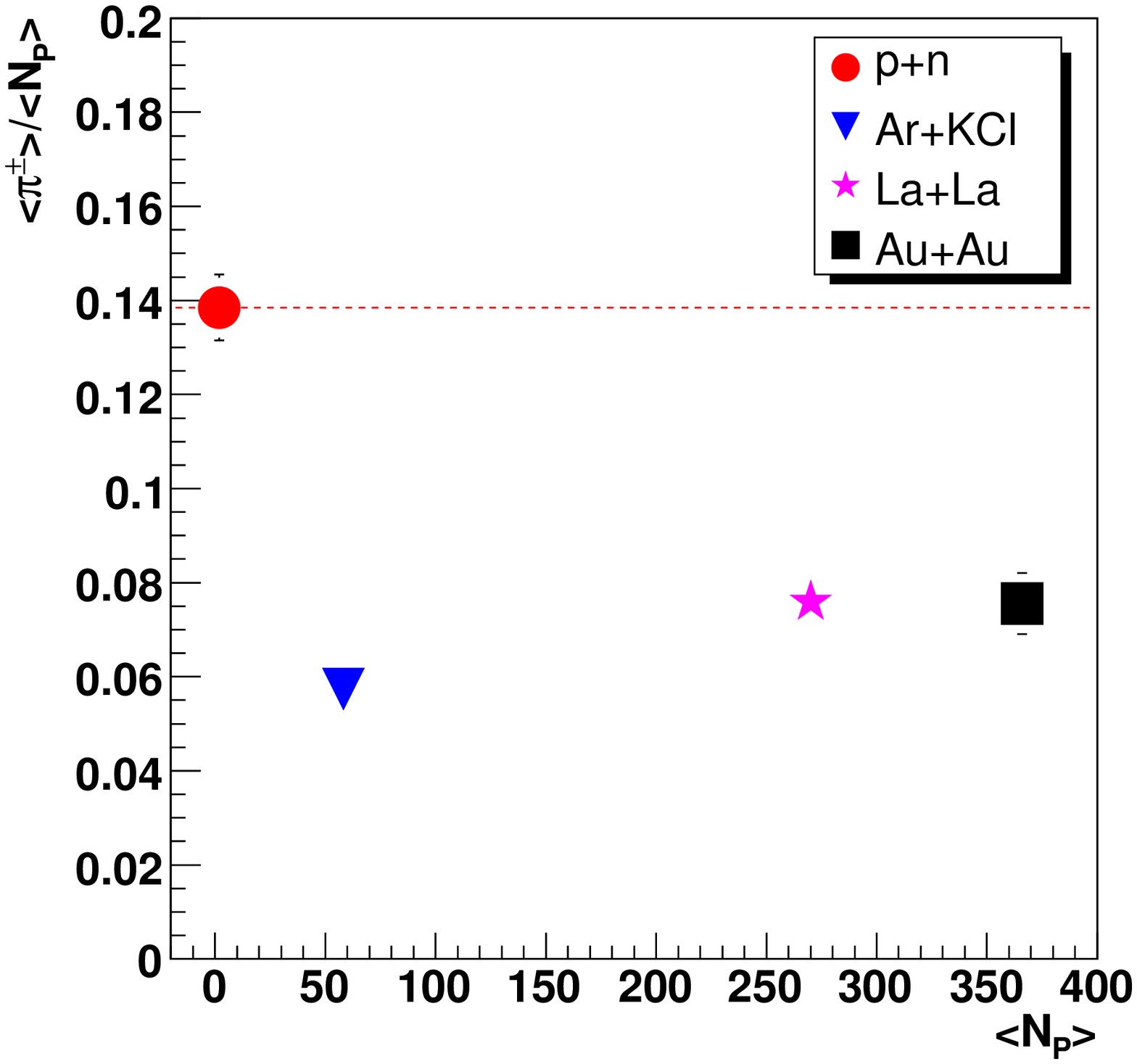}
\includegraphics[height=5cm]{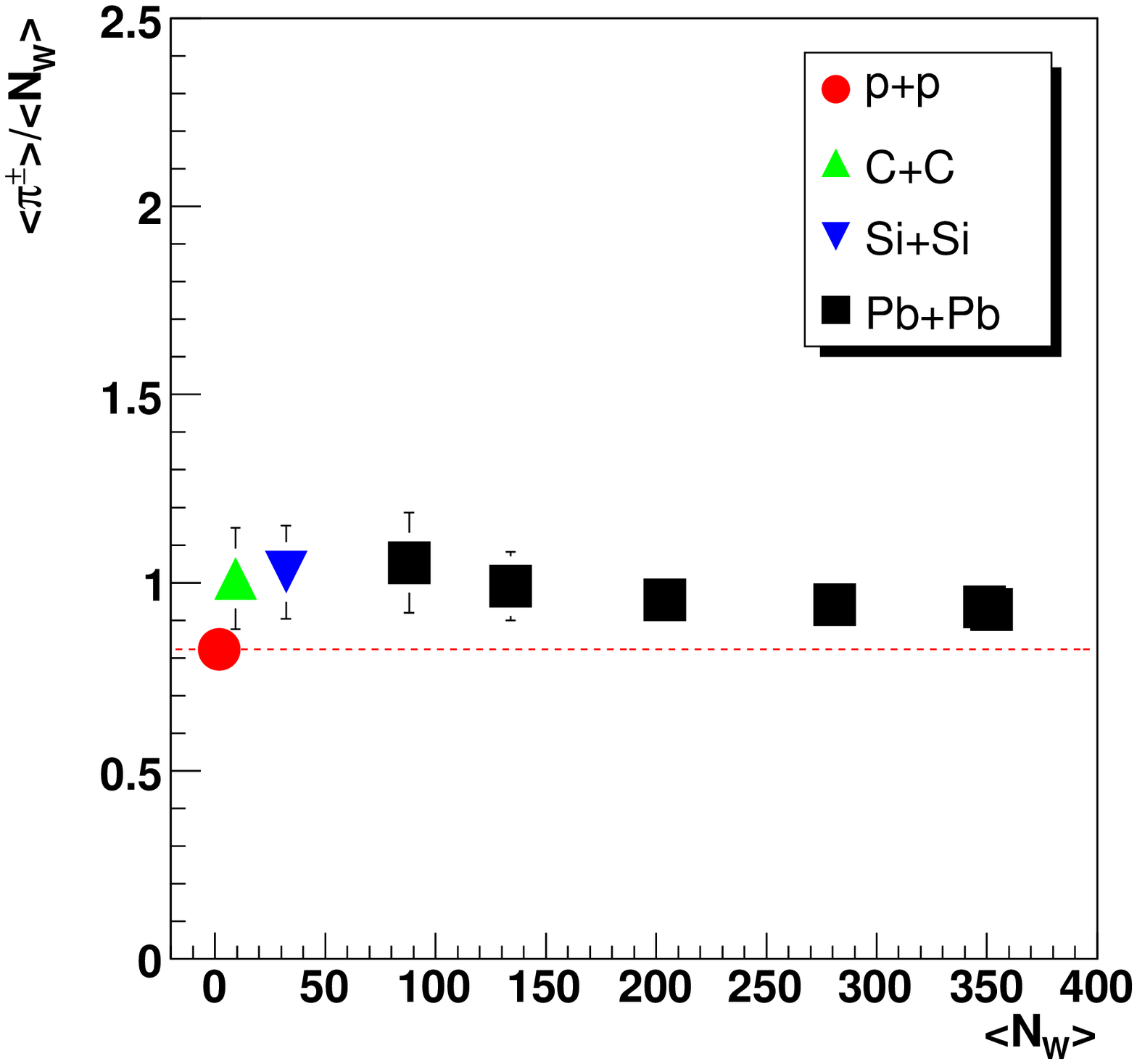}
\includegraphics[height=5cm]{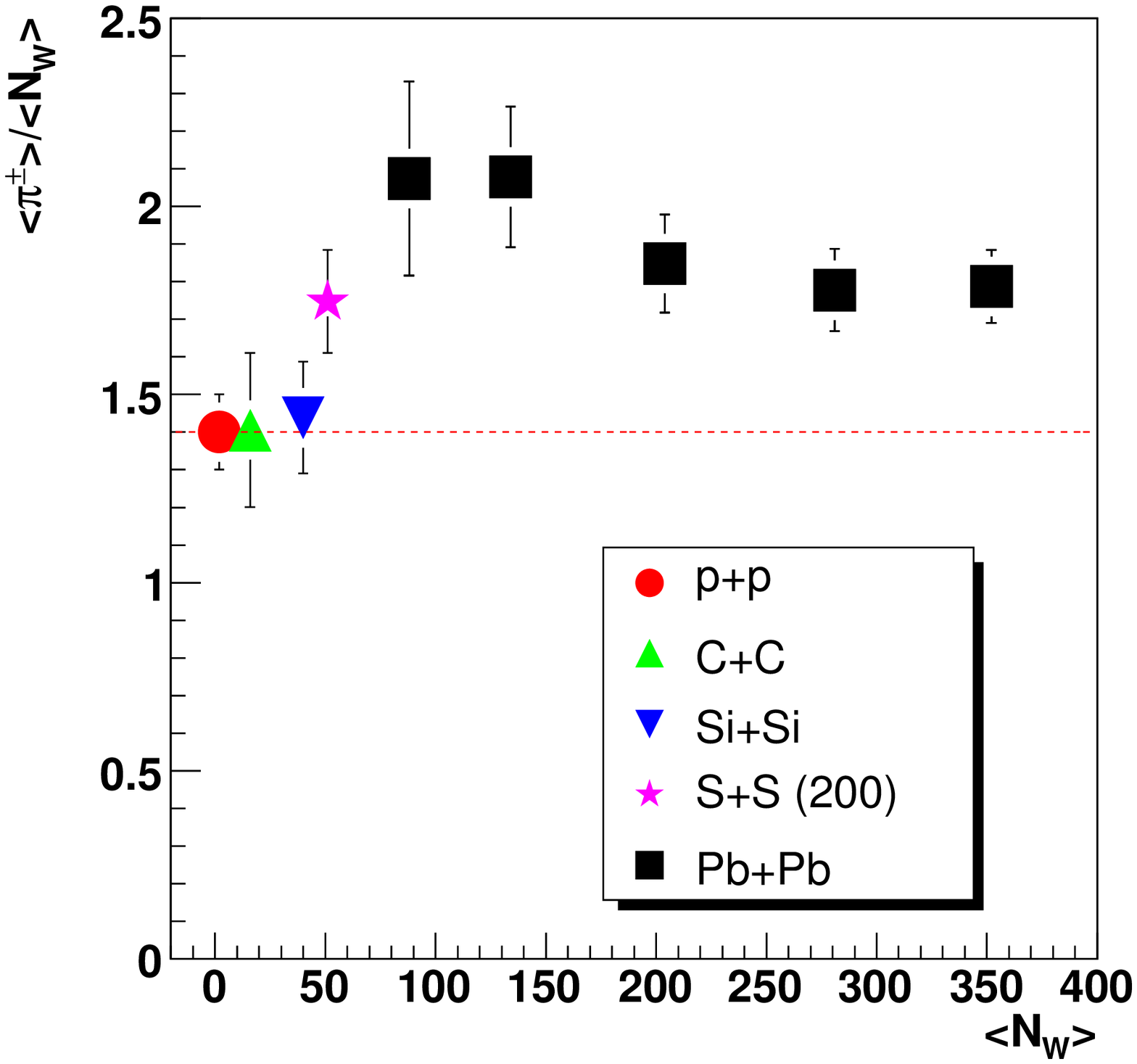}
\caption{\label{pi_ssd}System size dependence of pion yield for $2A$ GeV~\cite{Gazdzicki:1995zs, Klay:2003zf} (left), $40A$ GeV~\cite{Dinkelaker:2005py} 
(middle) and $158A$ GeV~\cite{Alt:2004wc,Bachler:1999hu} (right).}
\end{figure}

The system size dependence of pion production for various energies is shown in figure~\ref{pi_ssd}. Both the suppression at low energies and 
the enhancement at high energies start already in small systems.

%\subsection{Kaon production}
\vspace{0.4cm}
\noindent\textbf{Kaons:}\\
Strangeness is a very interesting observable because it is expected to be produced differently in a hadron gas and in the QGP. At SPS energies 
s-quarks are carried mainly
by $\Lambda$s and anti-kaons, while most $\rm{\bar{s}}$-quarks are carried by kaons. Therefore the $K^+$ meson multiplicity is approximately
equal to half to the
total number of  $\rm{\bar{s}}$ (and therefore also s) quarks produced in the collision. 
In order to take out the influence of center of mass energy on particle production the strange hadron yields are divided by the number of pions.

The energy dependence of the $\left<K^+\right> / \left<\pi^+\right>$ ratio shows (figure~\ref{pi_ed}) the famous "horn" which can not be reproduced by hadron gas models assuming $\gamma_S=1$
or string hadronic models. The increase of the $\left<K^-\right> / \left<\pi^-\right>$ ratio slows down at low SPS energies.
Both features are consistent with a model assuming a first order phase transition~\cite{Gazdzicki:1998vd} in which they are explained by 
%a reduction of the mass of strangeness carriers and
a lower fraction of strange to non-strange degrees of freedom in the QGP than in the hadron gas.
\begin{figure}[htp]
\includegraphics[height=5cm]{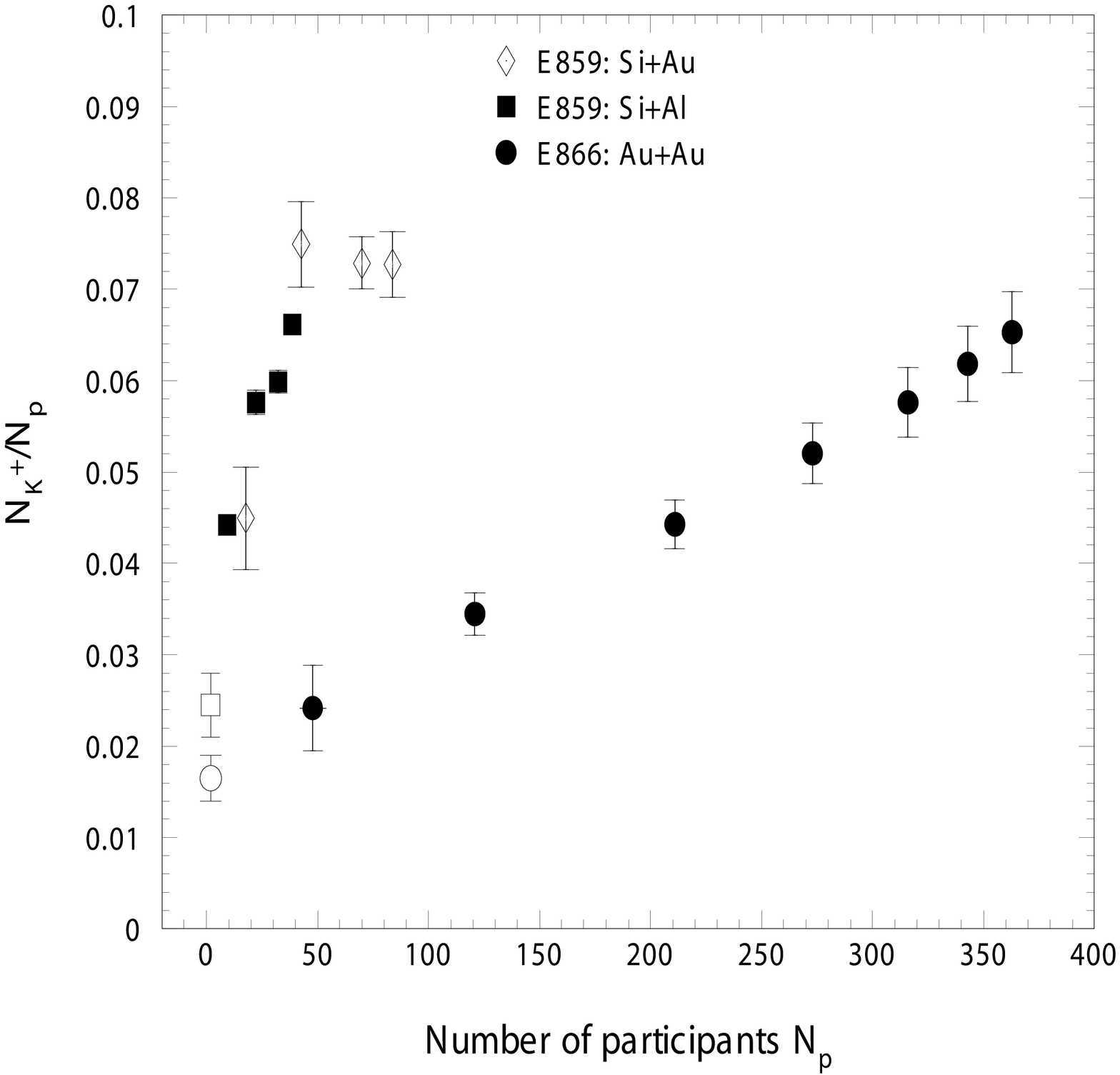}
\includegraphics[height=5cm]{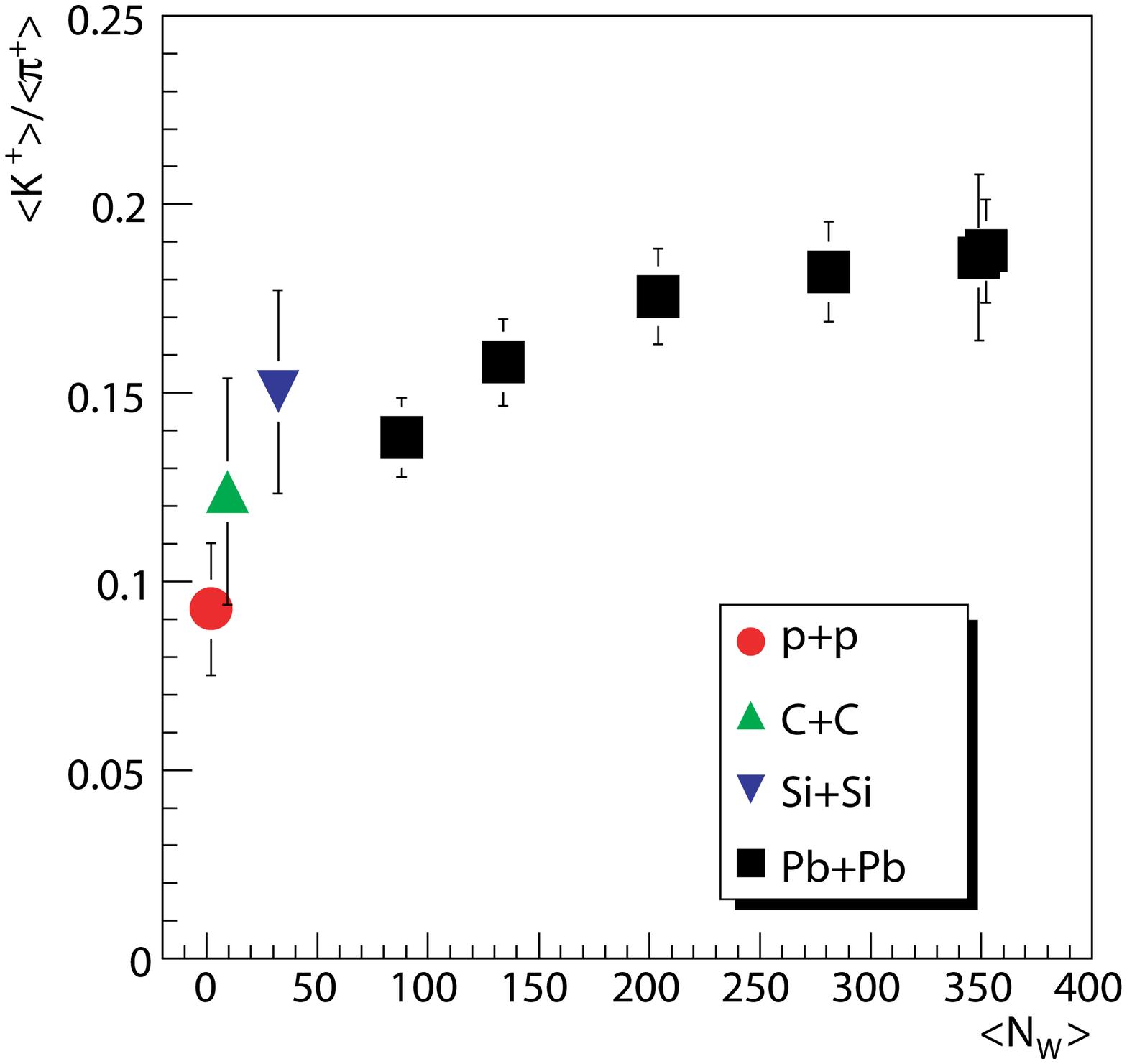}
\includegraphics[height=5cm]{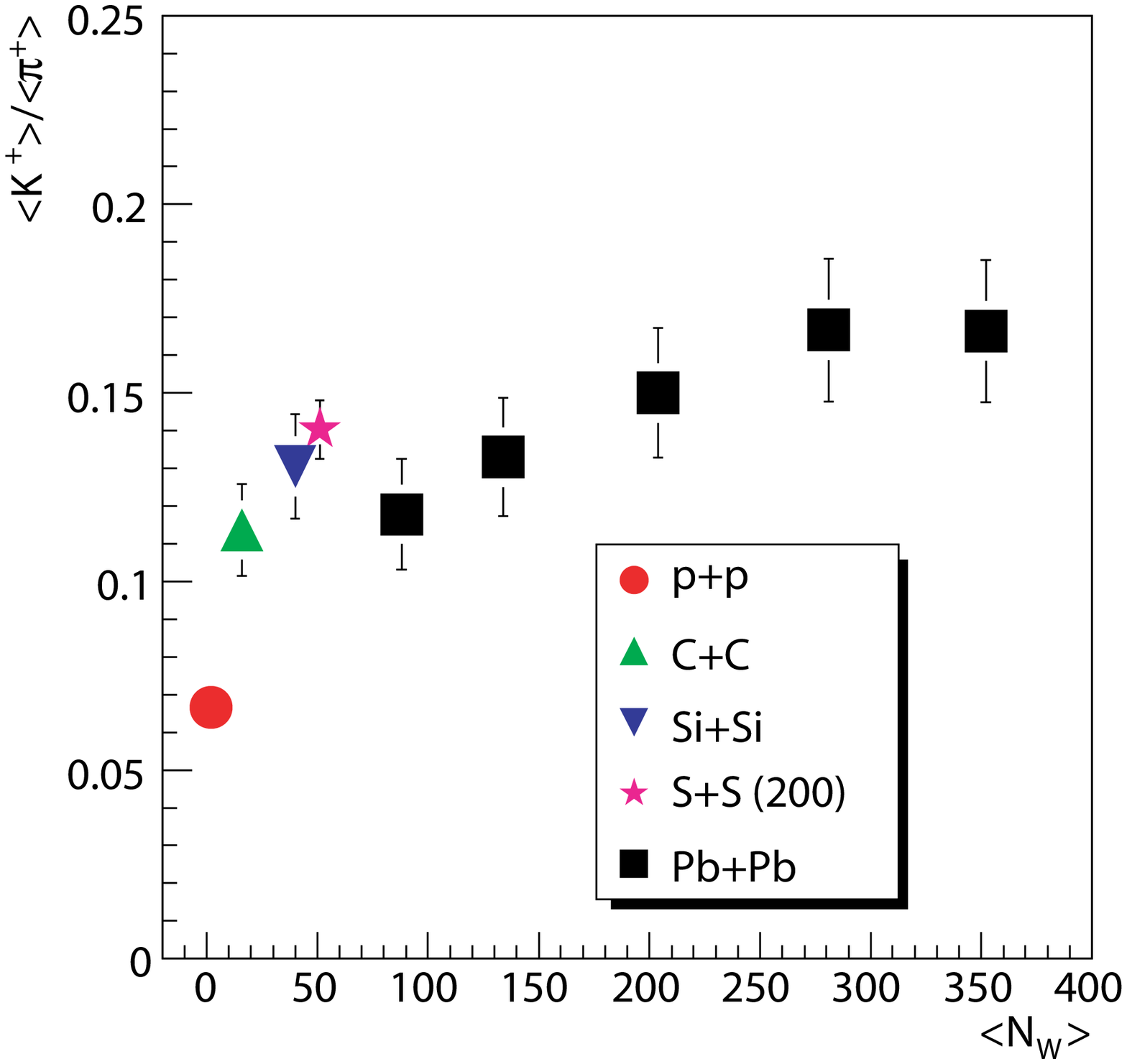}
\caption{\label{ssd_K}System size dependence of the $\left<K^+\right> / \left<\pi^+\right>$ ratio at $5A$ GeV~\cite{Wang:2000ak}, $40A$ GeV~\cite{Dinkelaker:2005py}
 and $158A$ GeV~\cite{Alt:2004wc,Bachler:1999hu}.}
\end{figure}

For central collisions the $\left<K^+\right> / \left<\pi^+\right>$ ratio rises quickly with system size for all
energies as shown in figure~\ref{ssd_K}. This effect can be attributed to the strong volume dependence of relative strangeness production, 
in statistical models known as
canonical strangeness suppression~\cite{Alt:2004wc,Hamieh:2000tk}.
%This is consistent with the picture of canonical suppression in small systems (ref ?).

The relative kaon production increases with centrality of the collision, the increase is faster and it saturates earlier for higher energies.

The number of wounded (or participant) nucleons $N_W$ (or $N_P$) is not a good scaling parameter.
The $\left<K^+\right> / \left<\pi^+\right>$ ratio for central collisions of small systems is larger, especially at lower energies, than the ratio for
peripheral collisions of large systems with the same number of wounded nucleons $\left<N_W\right>$.
This may be due to low collision density in peripheral reactions.

The $\left<K^-\right> / \left<\pi^-\right>$ ratio, not shown here, behaves qualitatively similar to the $\left<K^+\right> / \left<\pi^+\right>$ ratio.

%\subsection{Phi- meson production}
\vspace{0.4cm}
\noindent\textbf{$\phi$-mesons:}\\
The $\phi$ vector meson has zero net strangeness and is not expected to be correlated to total strangeness production if it is produced
according to the statistical hadron gas model.
However if the $\phi$ is produced by coalescence of a strange and an anti-strange quark its yield will be sensitive to the strange quark density
of the produced matter. 
\begin{figure}[htp]
\includegraphics[height=5cm]{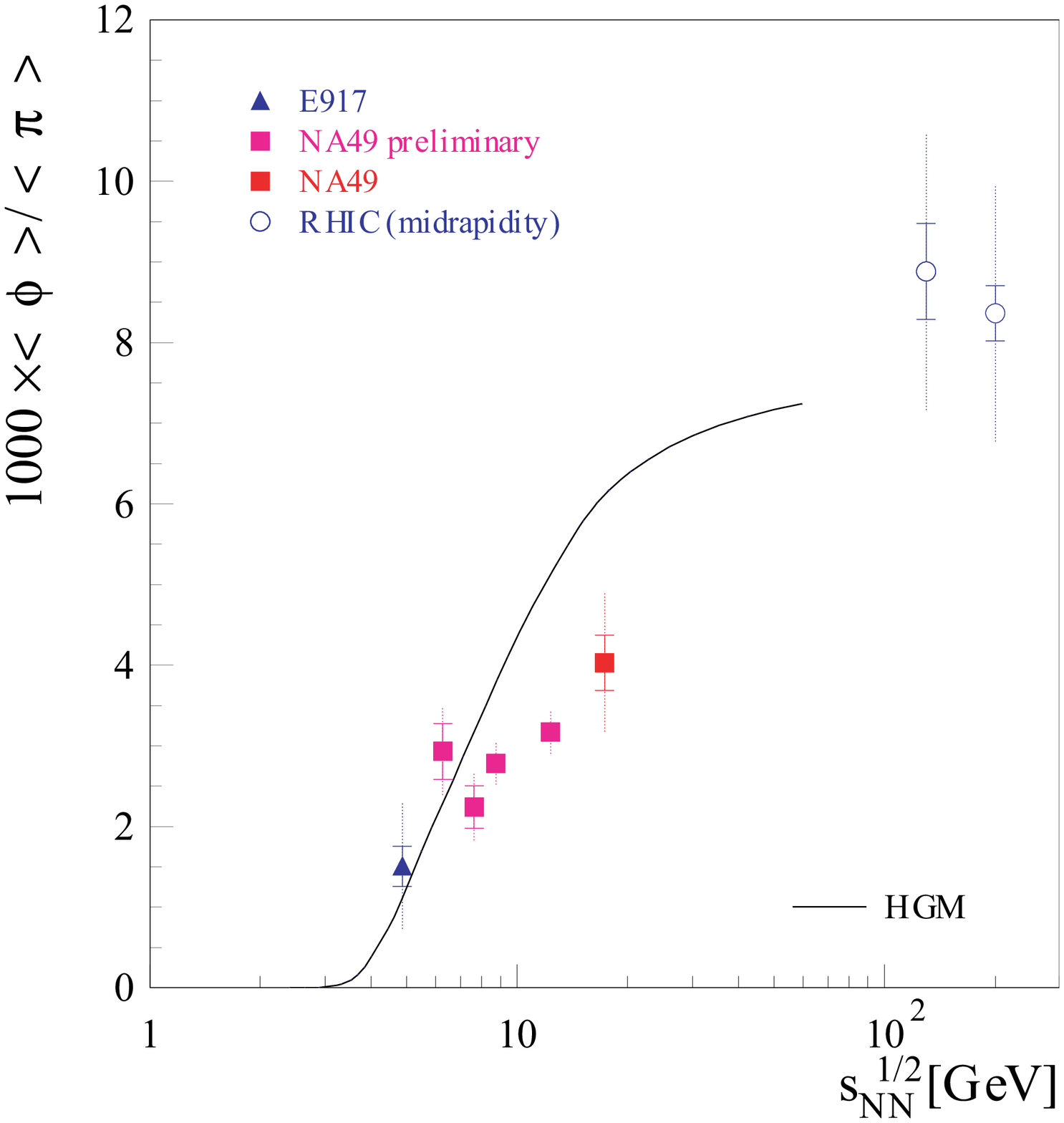}
\includegraphics[height=5cm]{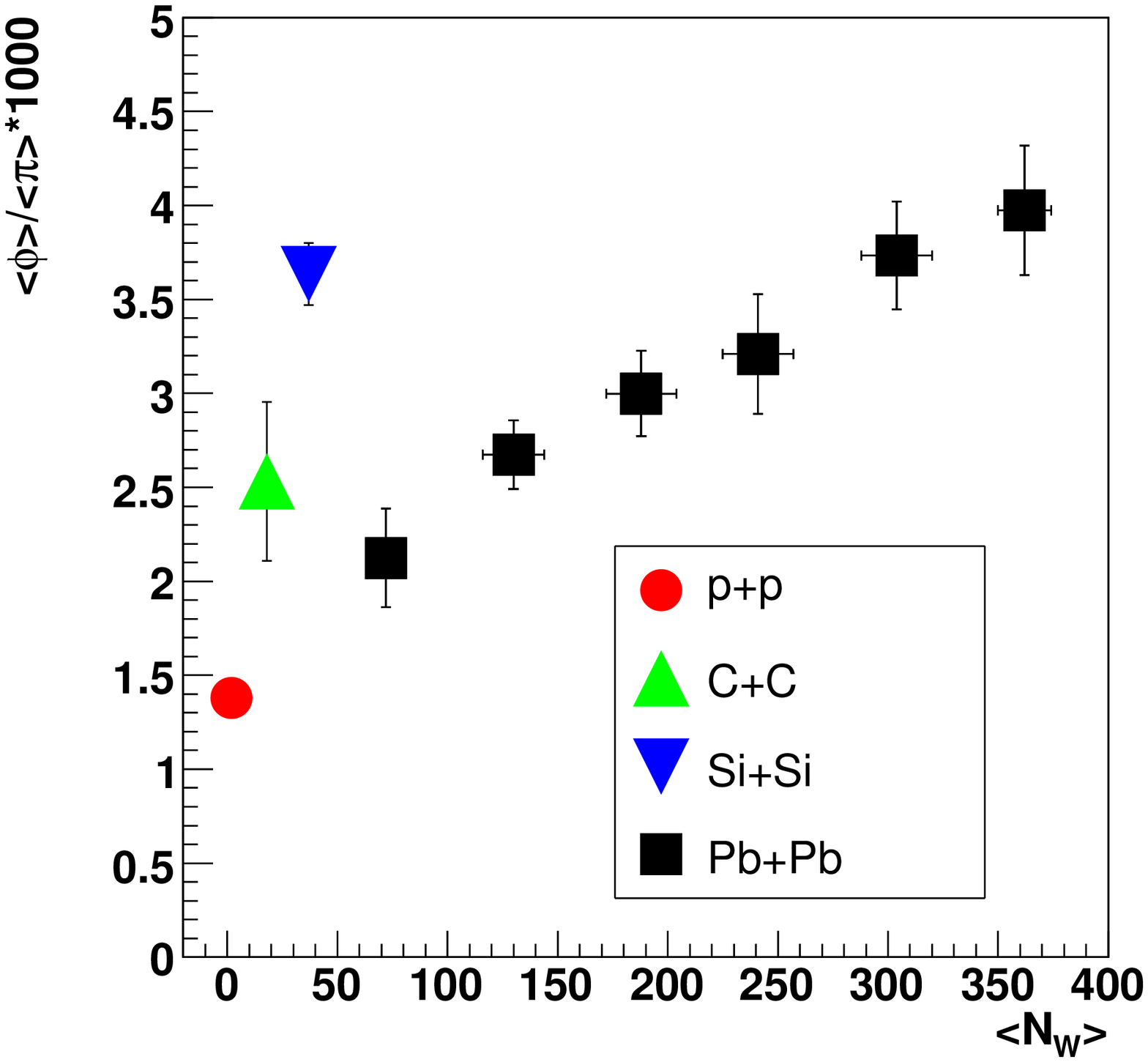}
\caption{\label{phi}Left: Energy dependence of the $\left<\phi\right> / \left<\pi\right>$ ratio for central Pb+Pb~(Au+Au) collisions~\cite{Afanasev:2000uu,Friese:2004av}.
Right: System size dependence of the $\left<\phi\right> / \left<\pi\right>$ ratio at $158A$ GeV~\cite{Alt:2004wc,Friese:2002re}.}
\end{figure}

The energy dependence of the $\left<\phi\right> / \left<\pi\right>$ ratio (figure~\ref{phi}) shows indications of a non-monotonic behavior at the low SPS energies. 
The system size dependence shows a similar structure as it was observed for the 
$\left<K^+\right> / \left<\pi^+\right>$ ratio. This indicates that $\phi$ production is sensitive to total strangeness production and favors the picture
of the $\phi$ production via coalescence of a strange and an anti-strange quark. 

\begin{figure}[htp]
\includegraphics[height=5cm]{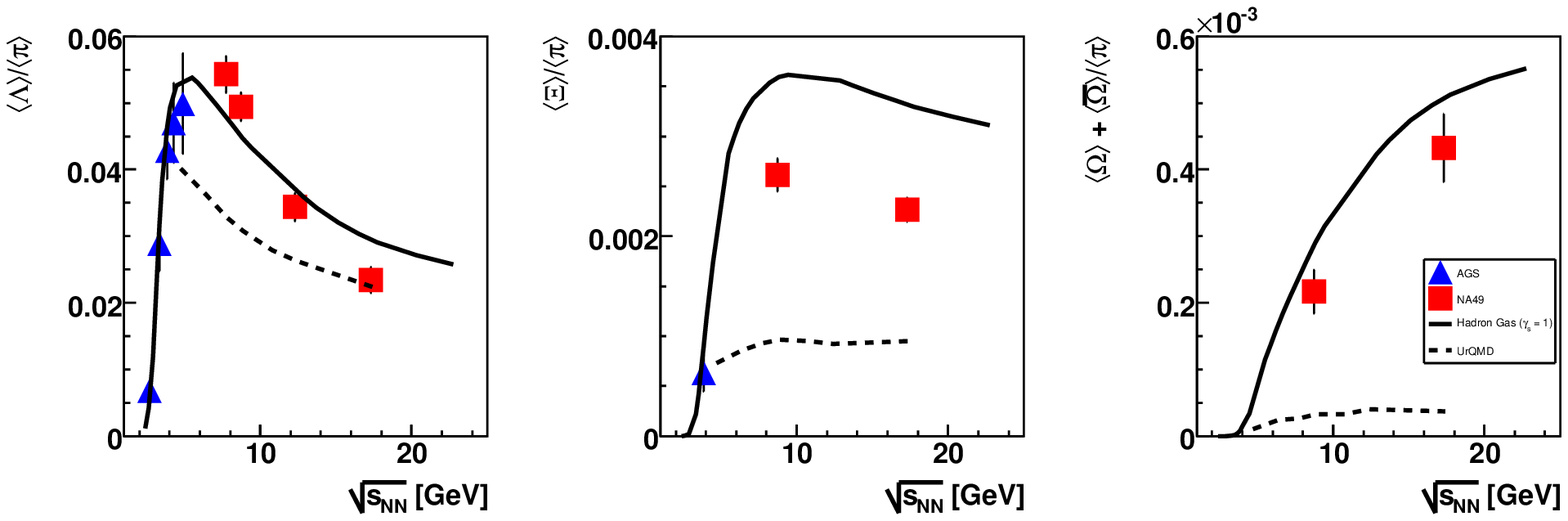}

\caption{\label{ed_hyp}Energy dependence of the $\left<\Lambda\right>/\left<\pi\right>$~\cite{Alt:2004wc,Friese:2004av}, 
$\left<\Xi\right>/\left<\pi\right>$~\cite{Friese:2004av} and
$\left<\Omega+\bar\Omega\right>/\left<\pi\right>$~\cite{Alt:2004kq} 
ratios in central Pb+Pb~(Au+Au) collisions.}
\end{figure}

%\subsection{Hyperon production}
\vspace{0.4cm}
\noindent\textbf{Hyperons:}\\
The energy dependence of the $\left<\Lambda\right> /\left<\pi\right>$ ratio shows a maximum at high AGS or low SPS energies. The position of that maximum is consistent
with the position of the maximum in the $\left<K^+\right> / \left<\pi^+\right>$ ratio. The energy dependence of $\left<\Lambda\right> /\left<\pi\right>$ can be approximately reproduced by a hadron gas 
model assuming
$\gamma_S=1$~\cite{Cleymans:1999st} (see solid curve in figure~\ref{ed_hyp}). 
The $\left<\Xi\right> / \left<\pi\right>$ ratio seem to have a maximum at low SPS energies like the $\left<K^+\right> / \left<\pi^+\right>$ and the $\left<\Lambda\right> /\left<\pi\right>$ ratios. 
The maximum is absent in the $\left<\Omega\right> / \left<\pi\right>$ ratio. The analysis of multi-strange hyperons is in progress and future results might clarify the situation.

Both the $\left<\Xi\right> / \left<\pi\right>$ and the $\left<\Omega\right> / \left<\pi\right>$ ratios at SPS energies are underestimated by the string hadronic UrQMD model~\cite{Bass:1998ca} 
(see curves in figure~\ref{ed_hyp}).

%\subsection{Total strangeness} 
\vspace{0.4cm}
\noindent\textbf{s and $\rm{\bar{s}}$ yield:}\\
The yields of $K^-$, $\Lambda$, $\Xi$ and $\Omega$ were used to calculate the total number of constituent s-quarks produced in the reaction.
The corresponding 
antiparticles were used to estimate the $\rm{\bar{s}}$-quark yield. The correction for the missing measurements was calculated based on the hadron gas 
model~\cite{Becattini:2003wp}.
\begin{figure}[htp]
\includegraphics[height=5cm]{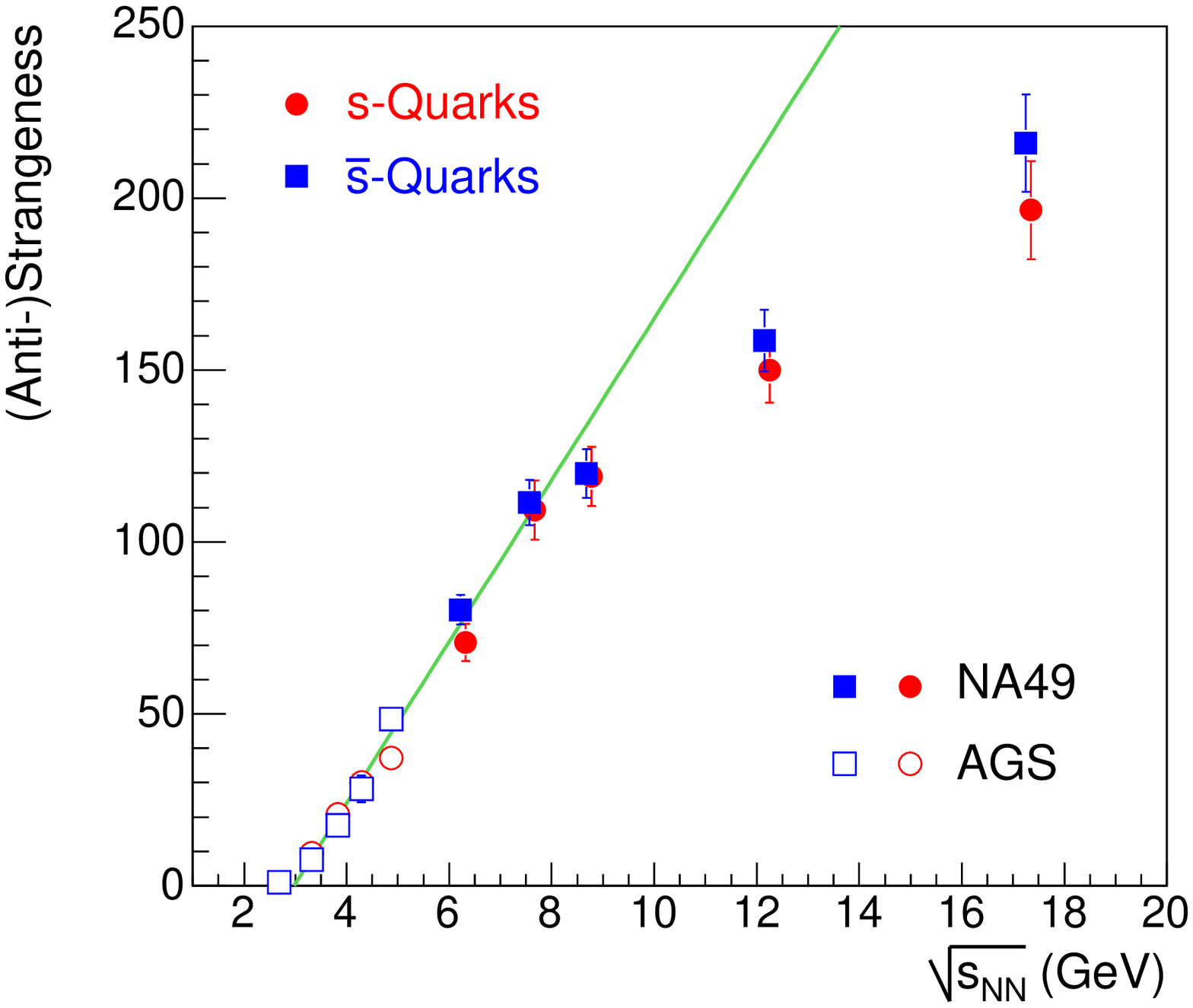}
\includegraphics[height=5cm]{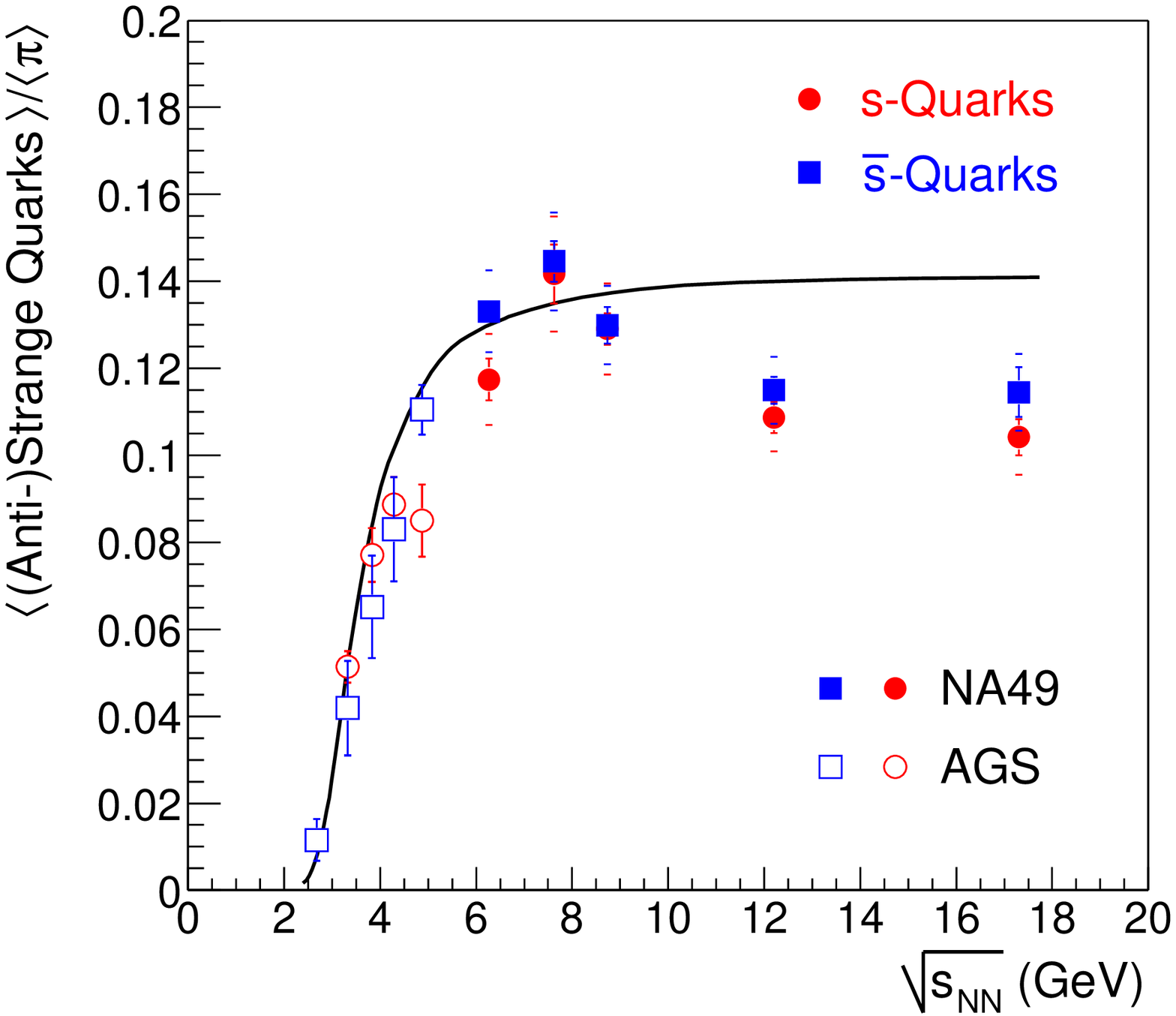}
\caption{\label{ed_str}Energy dependence of constituent s and $\rm{\bar{s}}$ quark multiplicities (left) and their ratios to the pion yield (right).}
\end{figure}

Figure~\ref{ed_str} shows that both NA49 and AGS data fulfill strangeness conservation $(\rm{\left<s\right>=\left<\bar{s}\right>})$. The "horn" which was observed in 
the $\left<K^+\right> / \left<\pi^+\right>$ ratio is
also seen in the $\left<s\right> / \left<\pi\right>$ and $\left<\bar{s}\right> / \left<\pi\right>$ ratios. The energy dependence of $\rm{\left<s\right>}$ and $\rm{\left<\bar{s}\right>}$
also shows an anomaly at low SPS energies: the increase of strangeness with energy is getting weaker at the low SPS energies. 

\vspace{0.4cm}
\noindent\textbf{Transverse mass spectra:}\\
The transverse mass spectra can be parametrized by the inverse slope parameter T obtained by fitting the function
$d^2 n/(m_T dy dm_T)=C \cdot \exp \left( -(m_T)/\rm{T} \right)$.
Collective flow effects cause a flattening in the low $m_T - m_0$  domain for heavier particles whereas at high $m_T - m_0$ the 
particle spectra follow rather a power law than an exponential dependence. Thus the fit range has to be limited to an intermediate $m_T$ range.
A simple exponential parametrization works best for kaons.

Another characterization of the shape of transverse mass spectra is the mean transverse mass $\left<m_T\right>$. Its advantage is that it can be used even for non-exponential
spectra.
\begin{figure}[htp]
\includegraphics[height=5cm]{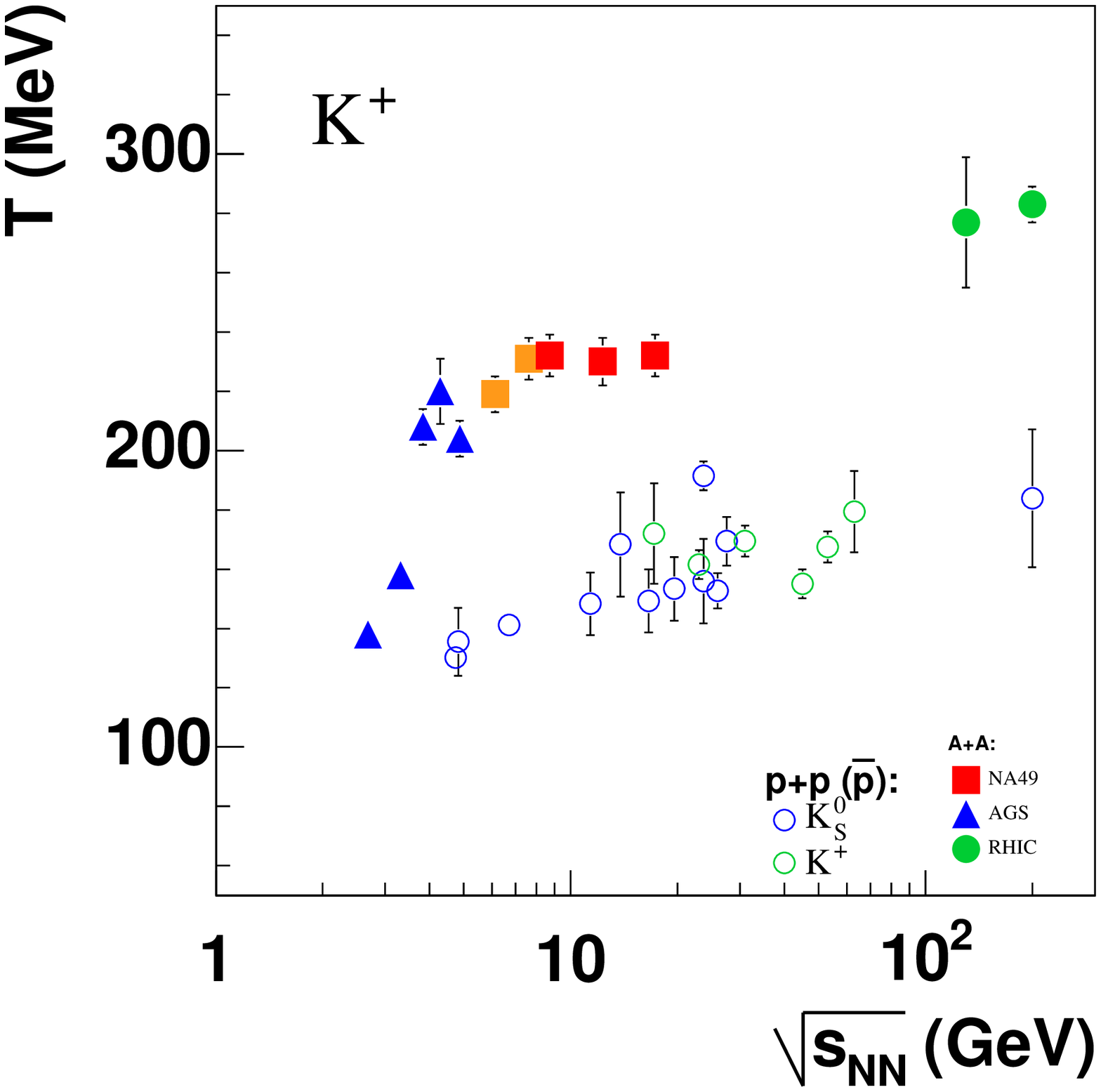}
\includegraphics[height=5cm]{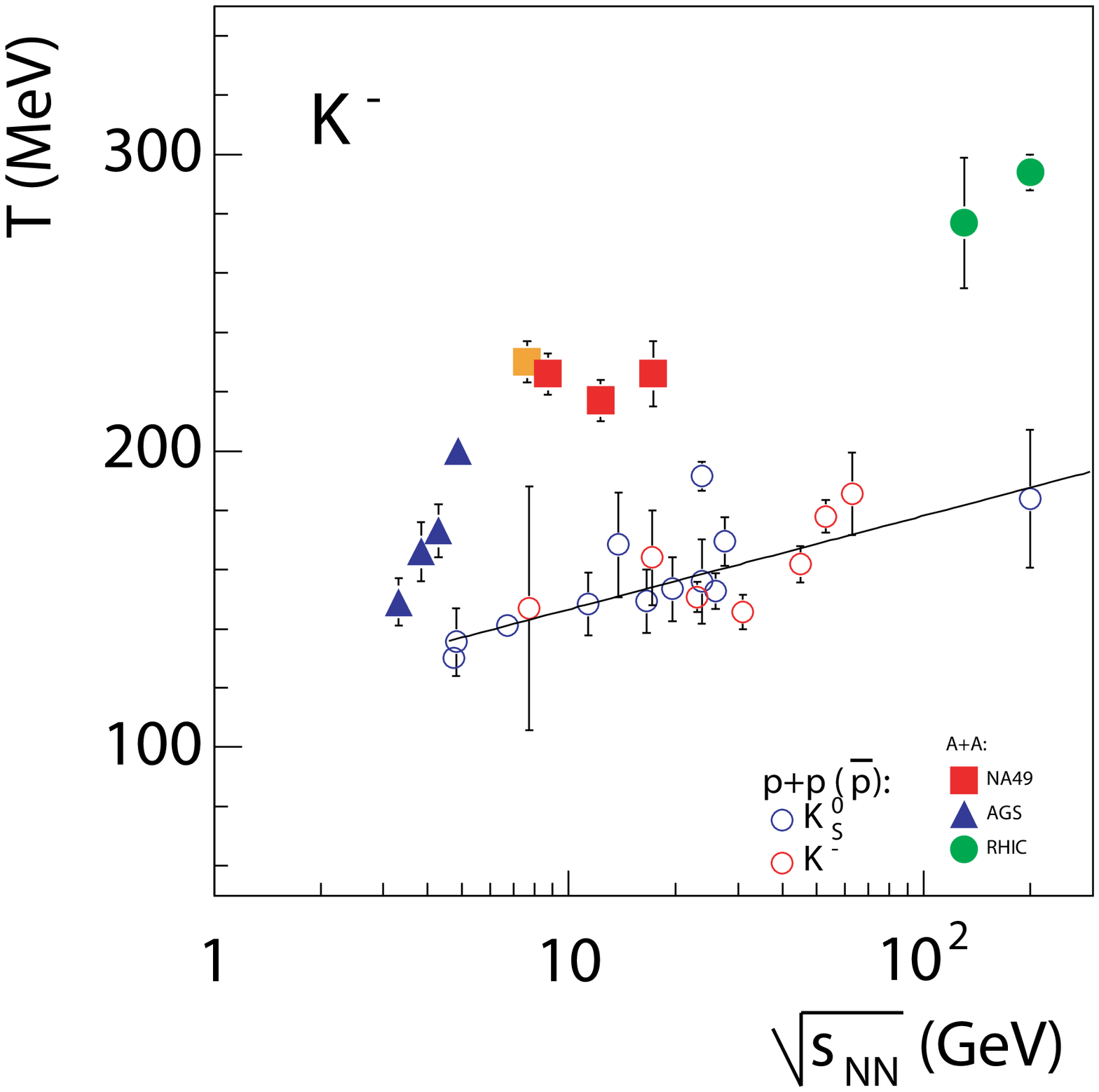}
\caption{\label{ed_KT}Energy dependence of the inverse slope parameter of kaons for central Pb+Pb~(Au+Au) and p+p~\cite{Kliemant:2003sa} collisions.}
\end{figure}

Figure \ref{ed_KT} shows that the inverse slope parameter of kaons increases in the AGS and RHIC energy domains but it stays constant at SPS energies. This feature, 
which is not seen in p+p interactions, might be attributed to the latent heat of a phase transition \cite{Gazdzicki:1998vd} and is in fact
consistent with hydrodynamic model calculations assuming a first order phase transition~\cite{Gazdzicki:2003dx}.
\begin{figure}[htp]
\includegraphics[height=6cm]{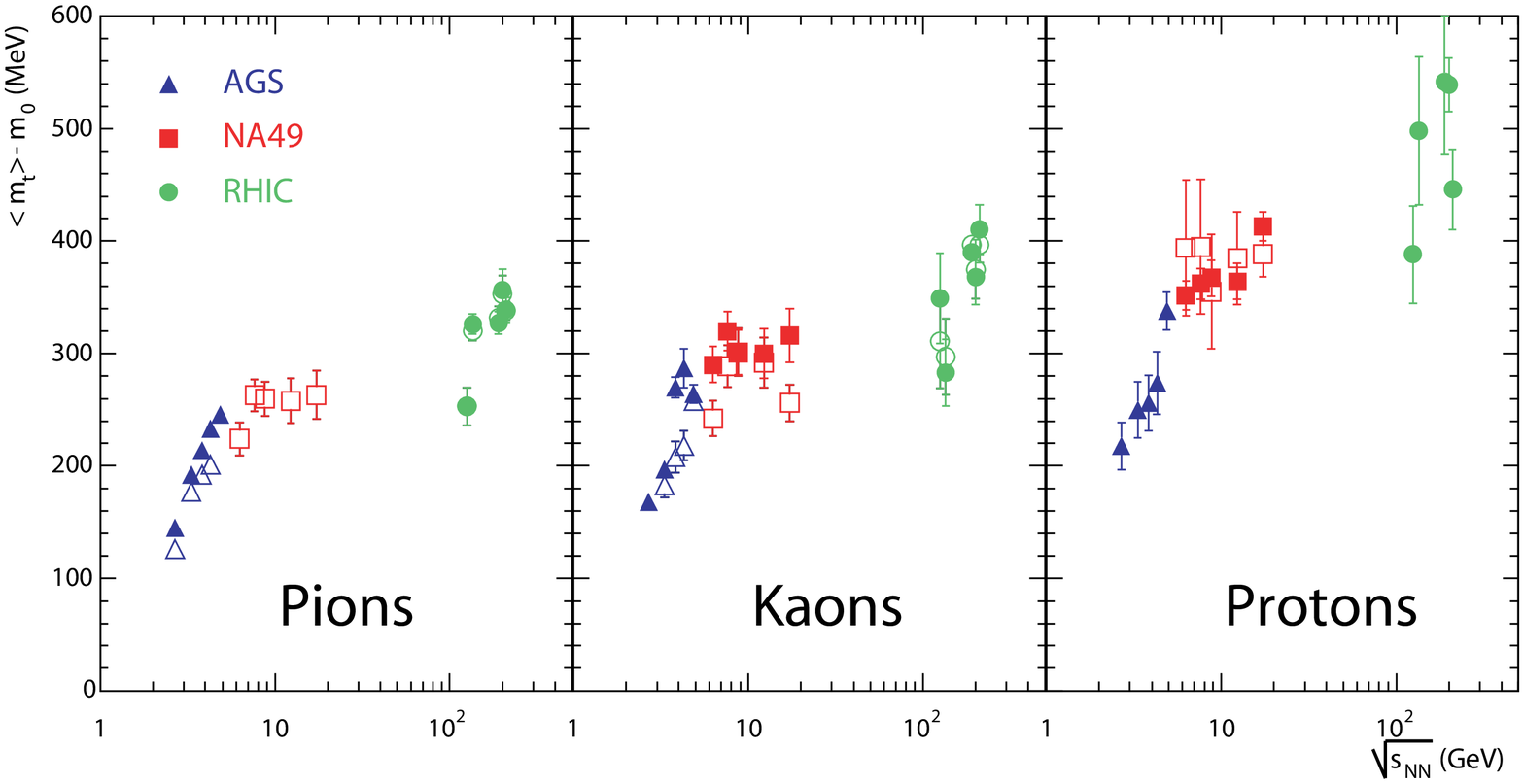}
\caption{\label{ed_mmt}Energy dependence of the mean transverse mass of various hadrons produced in central Pb+Pb~(Au+Au) collisions. Positive (negative)
particles are represented by full (open) symbols.}
\end{figure}

Similar energy dependence (the "step") is evident from figure~\ref{ed_mmt} for $\left<m_T\right>-m_0$. It seems to be present for pions, kaons and protons.

\begin{figure}[htp]
\includegraphics[height=5cm]{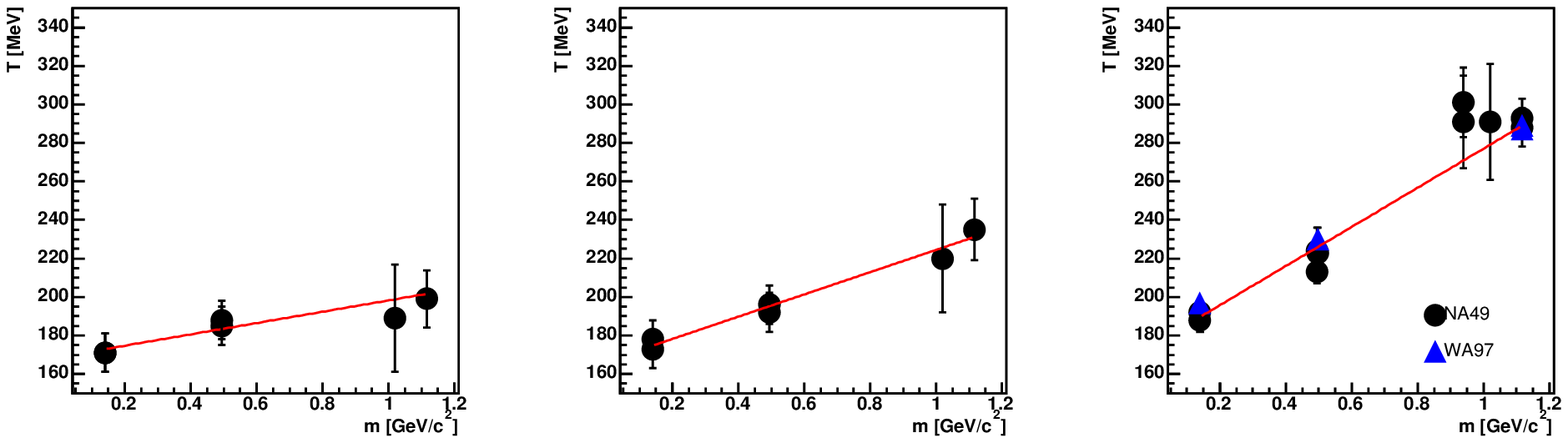}
\caption{\label{T_md}Mass dependence of the inverse slope parameter in central C+C, Si+Si~\cite{Alt:2004wc} 
and Pb+Pb~\cite{Gazdzicki:2004ef,Antinori:2000sb} 
collisions at $158A$ GeV.}
\end{figure}

The inverse slope parameter T fitted in the intermediate $m_T$ region rises linearly with particle mass with the slope being larger for larger systems 
(figure~\ref{T_md}).
This effect could be explained in the following way: The inverse slope has both a thermal component
and a contribution of the collective expansion. The latter gets larger for heavier particles and larger colliding systems.

\section{Conclusion}
Several anomalies in the energy dependence of hadron production properties were observed by the NA49 experiment. These are the "kink" for pion multiplicities,
the "horn" for the strangeness to pion ratio and the "step" for the mean transverse mass of various hadrons.
These observations can not be described by present hadronic models. In contrast, the results are consistent with the onset of deconfinement
at low SPS energies~\cite{Gazdzicki:1998vd,Letessier:2005qe}.

The system size dependence of various hadronic observables, e.g. the $\left<K\right>/\left<\pi\right>$ ratio, shows early saturation in central collisions of different
size nuclei. Peripheral
collisions of large nuclei behave quite differently from central collisions of small nuclei with the same number of wounded nucleons. This
difference is larger for smaller energies.

\vspace{0.4cm}
\noindent\textbf{Acknowledgements:} This work was supported by the US Department of Energy
Grant DE-FG03-97ER41020/A000,
the Bundesministerium fur Bildung und Forschung, Germany, 
the Virtual Institute VI-146 of Helmholtz Gemeinschaft, Germany,
the Polish State Committee for Scientific Research (1 P03B 097 29, 1 PO3B 121 29,  2 P03B 04123), 
the Hungarian Scientific Research Foundation (T032648, T032293, T043514),
the Hungarian National Science Foundation, OTKA, (F034707),
the Polish-German Foundation, the Korea Research Foundation Grant (KRF-2003-070-C00015) and the Bulgarian National Science Fund (Ph-09/05).

\bibliography{proceedings_lungwitz}

\end{document}